\documentclass[aps,pra,reprint,groupedaddress,showpacs,showkeys]{revtex4-1}
\usepackage{amsmath,graphicx,hyperref,commath}
\usepackage[FIGTOPCAP,raggedright,nooneline,bf,footnotesize]{subfigure}
\usepackage{indentfirst}
\usepackage{braket}
\usepackage{xcolor}
\usepackage{diagbox}
\usepackage{amssymb}
\usepackage{bbold}
\usepackage{multirow}

\newcommand{\tabref}[1]{Tab.~\ref{#1}}
\renewcommand{\eqref}[1]{Eq.~\ref{#1}}
\newcommand{\figureref}[1]{Fig.~\ref{#1}}

\newcommand{\tr}{{\mathrm{Tr}}}
\newcommand{\dd}{{\mathrm{d}}}
\newcommand{\centered}[1]{\begin{tabular}{l} #1 \end{tabular}}

\begin{document}

\title{Phase estimation of time-bin qudits by time-resolved single-photon counting}
\author{Artur Czerwinski}
\author{Karolina Sedziak-Kacprowicz}
\author{Piotr Kolenderski}
\email{kolenderski@fizyka.umk.pl}
\affiliation{Institute of Physics, Faculty of Physics, Astronomy and Informatics, Nicolaus Copernicus University, Grudziadzka 5, 87-100 Torun, Poland} 

\begin{abstract}
	We present a comprehensive framework for quantum state tomography (QST) of time-bin qudits sent through a fiber. Starting from basic assumptions, we define a positive-operator valued measure (POVM) which is then applied to the quantum state reconstruction problem. A realistic scenario is considered where the time uncertainty of the detector is treated as a source of experimental noise. The performance of the quantum tomography framework is examined through a series of numerical simulations conducted for different parameters describing the apparatus. The quality of state recovery, quantified by the notion of minimum fidelity, is depicted on graphs for a range of fiber lengths. Special attention is paid to relative phase reconstruction for qubits and qutrits. The results present relevant interdependence between the fiber length and the detector jitter.
\end{abstract}

\maketitle

\section{Introduction}

In quantum communication there is a demand for methods which provide complete characterization of quantum systems. The ability to manipulate well-characterized quantum resources is essential to encode information and plays a crucial role in quantum computing and cryptography. Since in general each measurement changes the state, we assume that the source can reproduce a large number of physical systems prepared in an identical quantum state. Thus, the problem of recovering the accurate representation of a quantum system relies on performing a series of measurements which provide complete information about the quantum state. This concept has been the subject of research since 1852, when G.~G.~Stokes developed the first technique to determine the polarization state of a light beam from intensity measurements \cite{Stokes1852}. Contemporary realizations of density matrix reconstruction based on polarization measurements were proposed by James \mbox{et al.~}\cite{James2001,Altepeter2004} and more recently by Bayraktar \mbox{et al.~}\cite{Bayraktar2016}.

For decades there have been different proposals of tomographic techniques, both theoretical and experimental. Some quantum state tomography (QST) methods aim to provide an explicit formula for an unknown density matrix based on data accessible from an experiment \cite{Abouraddy2002,Thew2002,Czerwinski2016}. However due to measurement imperfections, such approaches suffer from the risk of obtaining a matrix which does not belong to the state set and cannot describe a physical system. For this reason, in this article we follow numerical methods of state estimation which allow us to determine the density matrix which fits optimally to the experimental data. More specifically, we employ the maximum-likelihood estimation (MLE) \cite{Hradil1997} and the method of least squares (LS method) \cite{Opatrny1997}. Both approaches to QST have been commonly used to solve a variety of research problems, e.g. \cite{Banaszek1999,Altepeter2005,Smolin2012,Shang2017,SedziakKacprowicz2020}.

Quantum information can be encoded in photons by exploiting different degrees of freedom, especially: polarization, spectral, spatial and temporal mode. In this article, we follow the time-bin approach to quantum information encoding. Thus, for us a qubit is implemented by a photon delocalized in two wave packets separated in time. Initially, the concept of using temporal superposition of single photons was introduced in the context of Bell inequalities violation \cite{Franson1989}. Later, it was proposed as a resource for quantum communication \cite{Brendel1999,Marcikic2002}. Recent experiments \cite{Humphreys2013,Donohue2013}, which utilize time-bin encoded photons from an SPDC source, have demonstrated practical realizations of quantum information processing. Time-bin encoding allows one to overcome the problem of polarization-mode dispersion, which affects polarization encoded information. For this reason, the methods which facilitate the conversion of qubits between polarization and time-bin encoding are gaining in popularity \cite{Kupchak2017}. Since time-domain encoding is more resistant to thermal and mechanical disturbances, it has been successfully applied to long-distance quantum key distribution (QKD) \cite{Dynes2009,Inagaki2013,Ikuta2018} and quantum teleportation \cite{Valivarthi2016,Sun2016,Anderson2020}.

In this article, we combine the technique of time-bin encoding with the methods of QST. Section \ref{encoding} presents the necessary theory, which was originally presented in \cite{SedziakKacprowicz2020}. Starting from the definition of a qubit encoded in time domain, we can describe its dynamics when it travels through a fiber. Then, we define time-dependent measurement operators by employing the Heisenberg representation. Such a dynamic approach to quantum measurement can be applied to different types of quantum evolution in order to enhance QST, cf. \cite{Czerwinski2016a,Czerwinski2020}. The scenario is made real by imposing a detector timing jitter on the measured photon counts. Detection uncertainty is considered a limiting factor in quantum state engineering and, for this reason, theoretical models describing its formalism have been recently proposed \cite{Gouzien2018,Gouzien2019}. The assumption that the timing uncertainty is the source of experimental noise differentiates the present work from \cite{SedziakKacprowicz2020}, where the measured photon counts were distorted by the Poisson noise. In addition, contrary to \cite{SedziakKacprowicz2020}, here we focus on performing QST with the minimal number of measurements.

The main novel results are introduced in Section \ref{tomography}, which consists of four parts. First, we precisely describe the framework and provide all assumptions. Next, we define two figures of merit which can be used to evaluate the quality of QST. Equipped with this background, we perform numerical simulations and discuss the efficiency of state reconstruction for qubits and qutrits, taking into account different values of experimental parameters. Two kinds of the problem are distinguished -- the complete QST and the recovery of relative phase factors. The analysis is a follow-up of the works reported in Ref.~\cite{SedziakKacprowicz2020}. In the present work, we evaluate the efficiency of the tomographic technique for a wide range of parameters and the results are plotted on graphs. This work leads to valuable conclusions which indicate a potential for future experimental applications.

\section{Encoding and measuring a single qubit in the temporal mode}\label{encoding}

In the time-domain approach to quantum information encoding, a single qubit can be defined as a photon delocalized in two wave packets separated in time by an interval $\tau$. In order to characterize measurement, we study a propagation of a time-bin qubit through a fiber. Its state changes with respect to the length of the fiber $L$ and a dispersion parameter $\beta$ \cite{Sedziak2017}.

Then, we utilize the Born rule to get the probability density of photon detection at time $t$. The probability density can be associated with a measurement operator defined in the time domain:
\begin{equation}\label{operator1}
\hat{M} (t)  = \mu (t) \ket{\psi_M (t)} \bra{\psi_M (t)},
\end{equation}
where $\mu (t)$ can be interpreted as the weight relating to the normalized state $\ket{\psi_M(t)}$  \cite{SedziakKacprowicz2020}. The measurement operator defined in \eqref{operator1} can be considered a continuous positive-operator valued measure (POVM).

In a realistic scenario, the measured probabilities are distorted by noise and errors. For that reason, we take into account the detector's timing uncertainty, which may have a great impact on the accuracy of measurement in the case of time-bin photons. Thus, we consider measurement operators influenced by the detector jitter which are constructed as:
\begin{equation}\label{operator2}
\hat{M}_D (t) = \int \hat{M}(t') \:q_D(t-t') \, \dd t',
\end{equation}
where $q_D(t)$ is a Gaussian function describing detection uncertainty, i.e.
\begin{equation}
q_D(t)=\frac{\exp \left(-\frac{t^2}{2 \sigma_D^2}\right)}{\sqrt{2 \pi  \sigma_D^2}}
\end{equation}
and $\sigma_D$ stands for the timing jitter. This parameter plays a crucial role in our framework. In the case of superconducting nanowire single-photon detectors (SNSPDs), the timing jitter is around $25$ ps \cite{Divochiy2018,Shcheslavskiy2016} and for state-of-the-art detectors it can reach even $1$ ps \cite{Korzh2020}.

In the same way, one can define qutrits (then qudits) in the time domain and describe the probability distribution of photon detection by means of the operators corresponding both to the ideal circumstances (no jitter) and the realistic scenario which includes the timing jitter \cite{SedziakKacprowicz2020}. Since the measurements have a solid algebraic representation, it is justified to apply these operators in QST.

\section{Quantum state tomography}\label{tomography}

\subsection{Methods}

One can implement QST methods to reconstruct the initial state of the photon by making measurements on many copies of the system. For an unknown density matrix $\rho$, we utilize the Cholesky decomposition, which ensures that the reconstructed matrix belongs to the state set \cite{James2001,Altepeter2004}:
	\begin{equation}\label{eq:kwiat}
		\rho =\frac{ W^{\dag} W }{\tr\{W^{\dag} W\}},
	\end{equation}
where $W$ denotes a lower triangular complex matrix which depends on a set of real parameters $\mathcal{W} = (w_1, w_2, \dots)$. The set $\mathcal{W}$ consists of $4$ elements in the case of qubits and $9$ for qutrits. Thanks to this factorization, the problem of recovering an unknown state can be formulated in terms of estimation of the parameters $\mathcal{W}$. Numerical methods can be applied to determine the optimal reconstruction for a given set of data. In our article, we implement MLE and the LS method to characterize an unknown quantum state. Both techniques can be compared in terms of their efficiency \cite{Acharya2019}.

The measurement operators revised in the previous section can be used as a source of data for density matrix reconstruction since they are an approximate POVM. Mathematically, we follow the Born rule to describe the probabilities of single-photon detection. When a measurement is repeated for $\mathcal{N}$ photons, then from the experiment we can obtain the values of the photon count. Noise and errors are inherent in any kind of measurement and, for this reason, we need methods that produce adequate estimates of quantum states. In this article, we assume that the errors attributed to our measurement results are brought about by the detector jitter.

By $n_k^E$ we shall denote the expected photon count delivered by the $k-th$ measurement operator, whereas $n_k ^M$ shall refer to the photon count actually measured. The first figure can be computed by following the Born rule with the operators from \eqref{operator1}, i.e.
\begin{equation}\label{eq:expected}
n_k^E = \mathcal{N} \: \tr \{\hat{M} (t_k) \rho\}.
\end{equation}
In our scenario, the measured photon count is distorted due to the detector jitter. Therefore, the formula for $n_k ^M$ can be written as:
\begin{equation}\label{eq:measured}
n_k^M = \mathcal{N} \: \tr \{\hat{M}_D (t_k) \rho\}.
\end{equation}

By using the formula for the measurement operator including jitter we can model an imperfect scenario and simulate a set of experimental results distorted by the detector uncertainty. The simulated data is used to estimate an unknown quantum state based on MLE and LS method. In the case of MLE, we apply the likelihood function $\mathcal{L}$ in the form \cite{Ikuta2017}:
\begin{equation}
\mathcal{L}(\mathcal{W}) = \sum_{k} \left[\frac{(n_k^M -  n_k^E)^2}{ n_k^E }  + \ln n_k^E \right].
\end{equation}
The goal of MLE is to determine the parameters $\mathcal{W}$ which minimize the likelihood function. When it comes to LS, we follow the standard approach which allows us to approximate the solution by minimizing the sum of the squares of differences between the measured and expected photon counts.

As to compare the present work with \cite{SedziakKacprowicz2020}, it should be stressed that the photon counts \eqref{eq:expected}-\ref{eq:measured} are defined by means of different measurement operators. In \cite{SedziakKacprowicz2020}, both photon counts involved $\hat{M}_D (t)$ since the goal was to evaluate the performance of these imperfect operators in QST. However, the number of photons was distorted by the Poisson noise. In the current analysis, we assume that the number of photons per measurement is sufficiently large and the Poisson noise is negligible (thus, we have the same symbol $\mathcal{N}$ in $n_k^E$ and $n_k^M$).

Since the measurement operators are defined in the time domain, one needs to select a finite set of time instants, generating different operators. The moments have to cover a relevant range which corresponds to the weights $\mu (t)$ such that the measurements can be considered informative for state reconstruction. Natural questions concern the number of operators and the width of the time interval.

In QST, there is a strong tendency to search for methods which allow one to reduce the number of distinct measurements required for state recovery. For this reason, the first part of the research involved determining the optimal number of measurement operators. To do so, an initial number of operators was gradually reduced while the quality of state recovery was intently observed. It turned out that we can descend to $13$ operators without damaging the quality of quantum tomography. In \cite{SedziakKacprowicz2020}, QST was performed with $26$ measurement operators, which implies that we have reduced this number by $50\%$.

Therefore, hereafter in the study, each QST result is based on $13$ measurement operators. However, one needs to bear in mind that a light pulse spreads when it travels through a fiber. For this reason, the width of the time interval has to increase along with the length of the fiber. By a proper adjustment one can ensure that the time domain corresponds with the spectrum of the light pulse.

\subsection{Performance analysis}

In any realistic QST scheme, the reconstructed state $\rho_{out}$ differs from the input state $\rho_{in}$. It happens because the measurement results are always burdened with errors and noise. Therefore, we need to evaluate how well one can recover an unknown density matrix in spite of the detector jitter. In order to quantify the accuracy of our QST scheme, we introduce two figures of merit: \textit{the minimum fidelity} and \textit{the maximum trace distance}.

If one wants to compare the output state $\rho_{out}$ with the input state $\rho_{in}$, one can calculate the quantum fidelity from the formula \cite{Uhlmann1986,Jozsa1994}:
\begin{equation}\label{statesfidelity}
\mathcal{F}(\rho_{out}, \rho_{in}) = \left( \tr \sqrt{\sqrt{\rho_{out}}\rho_{in} \sqrt{\rho_{out}}} \right)^2.
\end{equation}
This figure is commonly used to describe how close the estimated state is to the actual one, e.g. Ref.~\cite{Jack2009,Nowierski2016,Oren2017}. Such a quantity can also be applied to test the efficiency of the quantum tomography framework in the case of simulated measurement results, e.g. Ref.~\cite{Yuan2016,Czerwinski2021}.

For a given measurement scheme, the value of quantum fidelity depends on the state $\rho_{in}$ introduced into the algorithm. However, we would like to have a single figure to evaluate the accuracy of our QST framework. Thus, we need to consider a discrete subset of input states, denoted by $\mathcal{S}$, which is a representative sample of the state set. For each input state $\rho_{in}$, we first generate the expected and measured photon counts \eqref{eq:expected}-\ref{eq:measured}. Then, we apply the quantum state estimation methods (MLE and LS) to recover the state. Finally, we compare the reconstructed state $\rho_{out}$ with the input $\rho_{in}$ in terms of the quantum fidelity \eqref{statesfidelity}. Since the work is based on numerical simulations, for each input state we obtain the fidelity, which describes how accurately it can be recovered by the framework. Ultimately, we define \textit{the minimum fidelity} as the lowest value for the considered set: 
\begin{equation}\label{minimumf}
\mathcal{F}_{min} := \min_{\rho_{in} \in \mathcal{S}} \mathcal{F}(\rho_{in},\rho_{out}),
\end{equation}
where $\mathcal{S}$ refers to the set of input states (the sample).

The other figure of merit is based on the trace distance between quantum states \cite{Nielsen2000}:
\begin{equation}
\mathcal{D} (\rho_{out}, \rho_{in}) = \frac{1}{2} \tr | \rho_{out} -  \rho_{in}|,
\end{equation}
where we follow standard notation $|A| \equiv \sqrt{A^* A}$. Here, in order to describe the quality of our scheme, we search for the maximum value of the trace distance within the same set of input states, i.e.:
\begin{equation}\label{maximumt}
\mathcal{D}_{max} := \max_{\rho_{in} \in \mathcal{S}} \mathcal{D} (\rho_{out}, \rho_{in}).
\end{equation}

The definitions of the minimum fidelity and the maximum trace distance mean that we want to determine the worst-case performance of our quantum tomography scheme over the sample of states. Such quantities can be useful to discuss the effectiveness of our POVM in quantum state reconstruction. Both figures of merit provide bounds on closeness of input and output quantum states. The minimum fidelity has been used to discuss the performance of state reconstruction in the scenario with imperfect measurements \cite{Rosset2012} as well as to measure the distortion of quantum states caused by a noisy channel \cite{Marshall2017}.

The quality of state reconstruction depends on the parameters describing our experimental setup. We shall treat two parameters as variables -- detector jitter $\sigma_D$ and length of the fiber $L$. For quantum tomography of both qubits and qutrits, we initially consider $6$ combinations of the parameters and calculate the figures of merit which are a starting point for further analysis.

\subsection{Quantum tomography of qubits}

In the case of qubits, the figures of merit have been gathered in \tabref{fig:qubit}. One can observe that if we assume that the scenario is perfect, i.e. there is no noise due to the jitter, then one can flawlessly reconstruct the initial state. This observation is rather intuitive since for $\sigma_D = 0$ ps the detector does not introduce any uncertainty into the measurements.

\begin{table}[h]
\centering
\bgroup
\def\arraystretch{2.1}
	\begin{tabular}{|c|c|c|c|c|c|}
\hline
		\multicolumn{2}{|c|}{\multirow{2}{*}{
				\backslashbox[22 mm]{\color{black}$\sigma_D$}{\color{black}$L\;$}}} & \multicolumn{2}{c|}{\centered{$200$ m}} & \multicolumn{2}{c|}{\centered{$500$ m}} \\ \cline{3-6} 
		\multicolumn{2}{|c|}{} & $\;\mathcal{F}_{min}\;$ & $\;\mathcal{D}_{max}\;$ & $\;\mathcal{F}_{min}\;$ & $\;\mathcal{D}_{max}\;$ \\ \hline
		\multirow{2}{*}{$0$ ps} &  LS  & $\; 1 \;$  &$\sim 10^{-6}$  & $ 1 $ &$\sim 10^{-6}$ \\ \cline{2-6} 
		& MLE & $1$ & $\sim 10^{-6}$ & $1$ & $\sim 10^{-6}$    \\ \hline

		\multirow{2}{*}{$1$ ps} & LS & $0.7803$ & $0.2222$ & $0.9542$ & $0.0459$  \\ \cline{2-6} 
		& MLE & $0.7744$ & $0.2259$ & $0.9532$ & $0.0472$ \\ \hline

		\multirow{2}{*}{$4$ ps} & LS & $0.4456$ & $0.5544$ & $0.6152$ & $0.3849$  \\ \cline{2-6} 
		& MLE & $0.4844$ & $0.5156$ & $0.6105$ & $0.3909$ \\
\hline
	\end{tabular}
\egroup
	\caption{Minimum fidelity and maximum trace distance in quantum tomography of qubits. Figures computed numerically for different values of experimental parameters. The results were obtained after minimization over a sample of $9261$ qubits.}
	\label{fig:qubit}
\end{table}

For $L=200$ m one can notice that when the timing jitter increases, the state reconstruction gets less accurate since the noise associated with the jitter distorts the measured counts more significantly. We observe that the maximum trace distance grows along with the timing jitter, whereas the minimum fidelity declines. Interestingly, both QST methods lead to very similar results.

The most interesting conclusion can be made if we compare the results for different lengths of the fiber. Both figures of merit change significantly if, for a nonzero jitter, we substitute the length $L=200$ m for $L=500$ m. One can observe that the quality of state reconstruction improves if we consider the longer fiber. In particular, for $\sigma_D = 1$ ps we can achieve a great deal of enhancement. This value of the detector jitter is experimentally achievable \cite{Korzh2020} and, therefore, these figures seem to have potential for future applications. This outcome suggests that the length of the fiber can be used to overcome the noise caused by the detector jitter.

\begin{figure}[h!]
	\centering
	\begin{tabular}{c}
		\centered{\includegraphics[width=1\columnwidth]{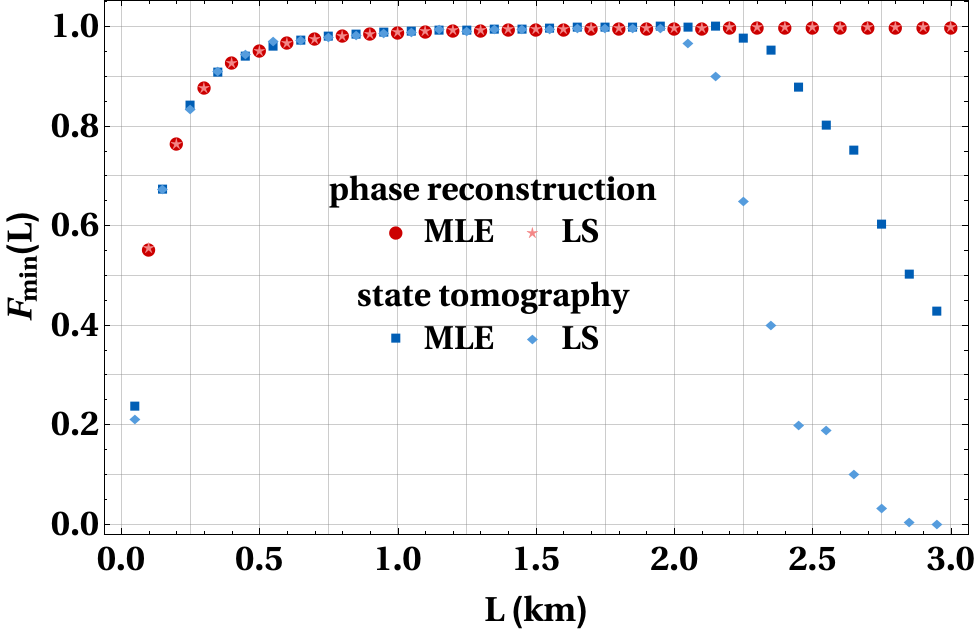}}
	\end{tabular}
	\caption{Plots of $\mathcal{F}_{min} (L)$ for complete state tomography and phase reconstruction of qubits by both QST methods. Detector jitter is fixed $\sigma_D = 1$ ps.}
	\label{plotsqubit}
\end{figure}

In order to analyze in detail the relation between the length of the fiber and the quality of state reconstruction, we need to consider the minimum fidelity as a function of $L$. In \figureref{plotsqubit}, one can see the results obtained for $\sigma_D = 1$ ps with $L$ ranging from $50$ m to $3000$ m by both QST methods (with step $100$ m).

The results in \figureref{plotsqubit} allow one to observe that the function $\mathcal{F}_{min} (L)$ for QST for complete state tomography is composed of three segments. First, $\mathcal{F}_{min} (L)$ increases along with the length of the fiber. Approximately for $L=1000$ m the function reaches its maximum value $\sim 0.99$ (according to both QST methods) and then remains roughly constant up to $L=2000$ m (LS method) or $L=2300$ m (MLE), when it starts to decline. One can observe that the performance of LS method degenerates more rapidly than in the case of MLE. For $L=3000$ m, the minimum fidelity by the LS method is close to zero, whereas in the case of MLE it is approximately $1/3$.

The declining character of $\mathcal{F}_{min} (L)$ is associated with the fact that for longer fibers quantum states corresponding to the measurement operators $\hat{M} (t)$ approach the equator of the Bloch sphere, which was shown in Ref.~\cite{SedziakKacprowicz2020}. However, within the Bloch ball there is a subset of quantum states which appears to be more resistant to this deterioration of the POVM's quality. As a special case, we shall consider a one-parameter family of input states given by:
\begin{equation}\label{input2}
\rho_{in} (\phi) = \frac{1}{2} \begin{pmatrix}  1 & e^{- i \,\phi} \\ e^{ i \,\phi} & 1  \end{pmatrix}.
\end{equation}
The input states of this form lie on the equator of the Bloch sphere and differ only in the relative phase $\phi$. However, when recovering the quantum state of the system, we assume that there is no \textit{a priori} knowledge about the state, which means that we still follow the formula given in \eqref{eq:kwiat} for $\rho_{out}$ and estimate all $4$ parameters characterizing a qubit density matrix. The graphs of $\mathcal{F}_{min} (L)$ for input states given by \eqref{input2} are also depicted in \figureref{plotsqubit} by different colors (each value of $\mathcal{F}_{min} (L)$ was computed after minimization over a sample of $201$ input states parametrized as in \eqref{input2}).

It turns out that when we admit only the states from the equator of the Bloch sphere, $\mathcal{F}_{min} (L)$ increases asymptotically to $1$ and then the quality of state reconstruction remains stable. The simulations have been carried on till $L=1\,000$ km and no deterioration was found out.

\begin{figure}[h!]
	\centering
	\begin{tabular}{c}
		\centered{\includegraphics[width=0.95\columnwidth]{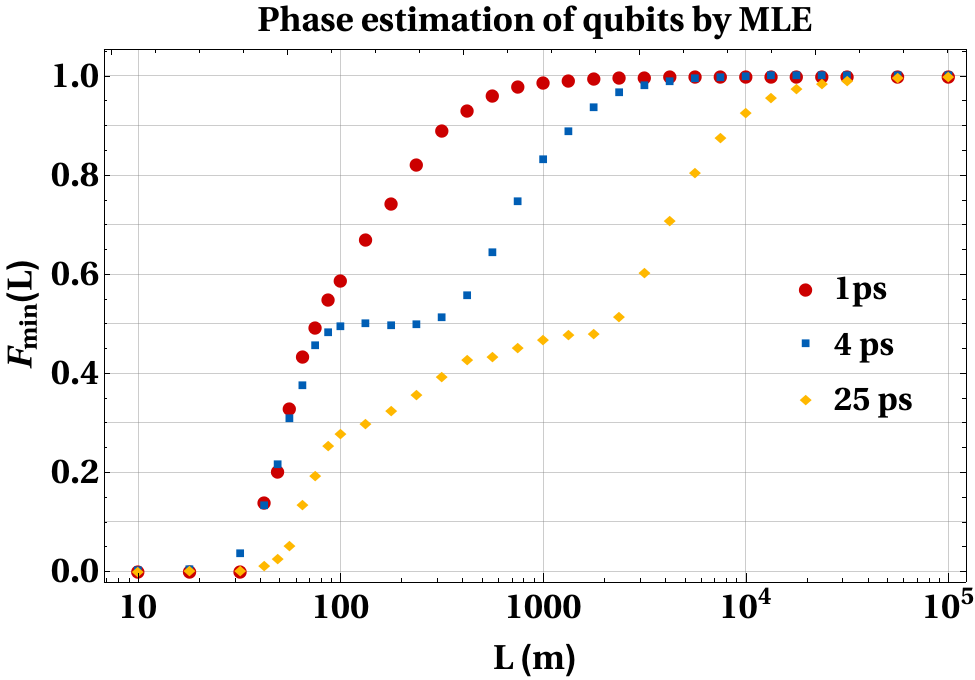}}
	\end{tabular}
	\caption{Plots of $\mathcal{F}_{min} (L)$ for phase estimation of qubits performed by MLE in the case of $\sigma_D = 1$ ps (red), $\sigma_D = 4$ ps (blue) and $\sigma_D = 25$ ps (yellow). Fiber lengths are arranged in a logarithmic scale.}
	\label{plotqubitphase}
\end{figure}

The quality of phase estimation can also be studied for other values of the timing jitter. Since $\sigma_D = 1$ ps corresponds with state-of-the-art technology, it would be desirable to investigate and compare how efficiently one can recover the relative phase with less sophisticated devices. In \figureref{plotqubitphase}, one can find the plots of $\mathcal{F}_{min} (L)$ in phase estimation for $\sigma_D = 1$ ps and two larger values of the detector jitter. We restrict the analysis only to MLE because there is no substantial difference between both tomographic techniques as far as the phase estimation is concerned.

For both larger values of the detector jitter, one can notice that the quality of phase estimation is steadily rising towards $1$. If $\sigma_D = 25$ ps, then $\mathcal{F}_{min} (L)$ exceeds $0.9$ when $L = 10^4$ m. On average, one obtains better quality of phase estimation if $\sigma_D = 4$ ps, which is not a surprise. However, the three functions converge for greater lengths of the fiber. To conclude, one may say that even for a realistic timing jitter the QST framework is still effective when it comes to the relative phase estimation provided one employs a fiber of sufficient length.

	\begin{figure}[h!]
		\centering
		\begin{tabular}{c|c}
			\backslashbox[12mm]{\color{black}$L$}{\color{black}$\sigma_D$}&\centered{$4$ ps}\\\hline
			\centered{$100$ m}&\centered{\includegraphics[width=0.45\columnwidth]{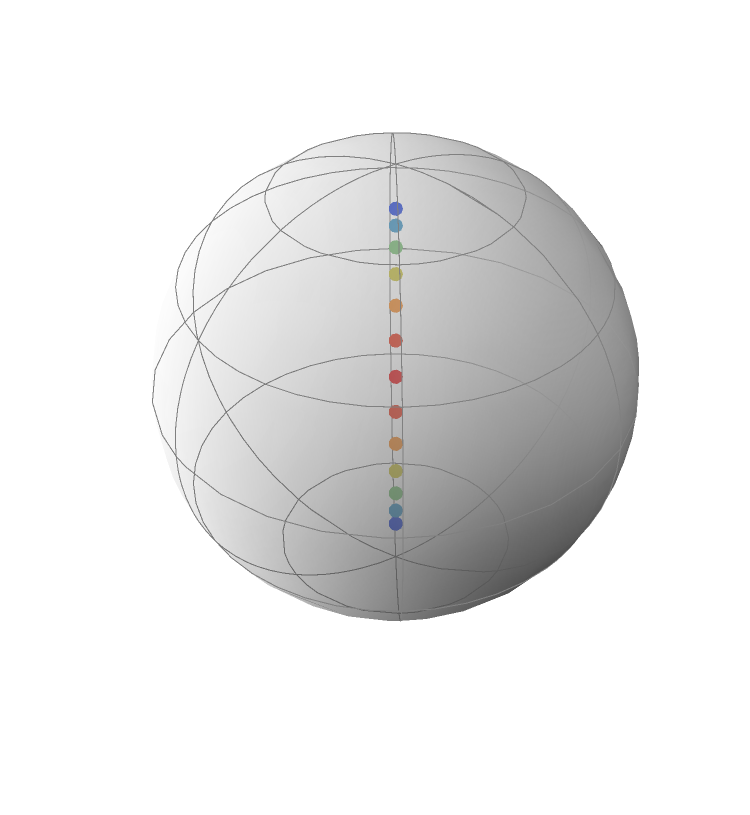}} \\ \hline
			\centered{$300$ m}&\centered{\includegraphics[width=0.45\columnwidth]{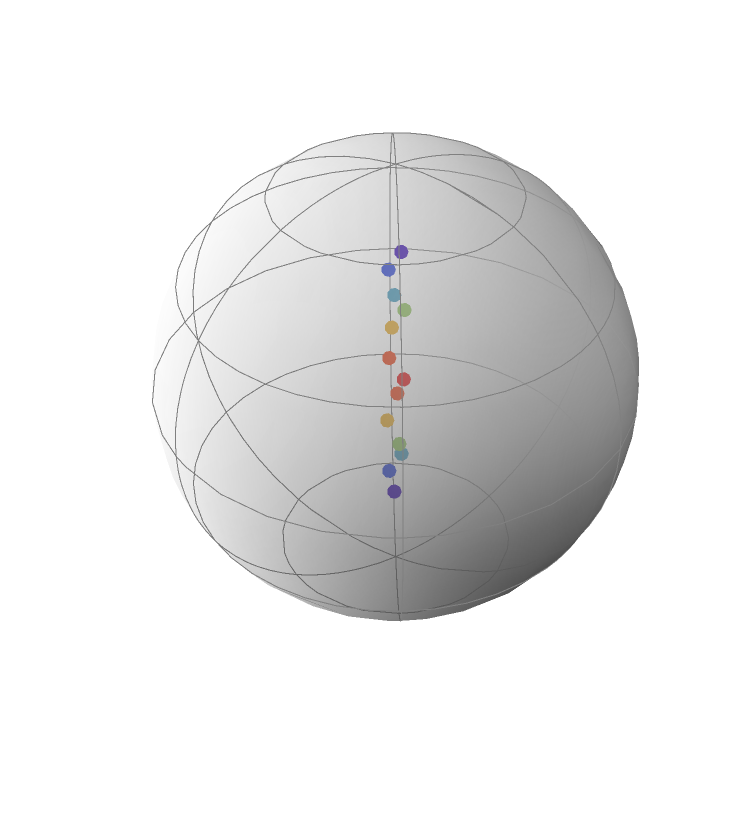}}\\ \hline
			\centered{$1000$ m}&\centered{\includegraphics[width=0.45\columnwidth]{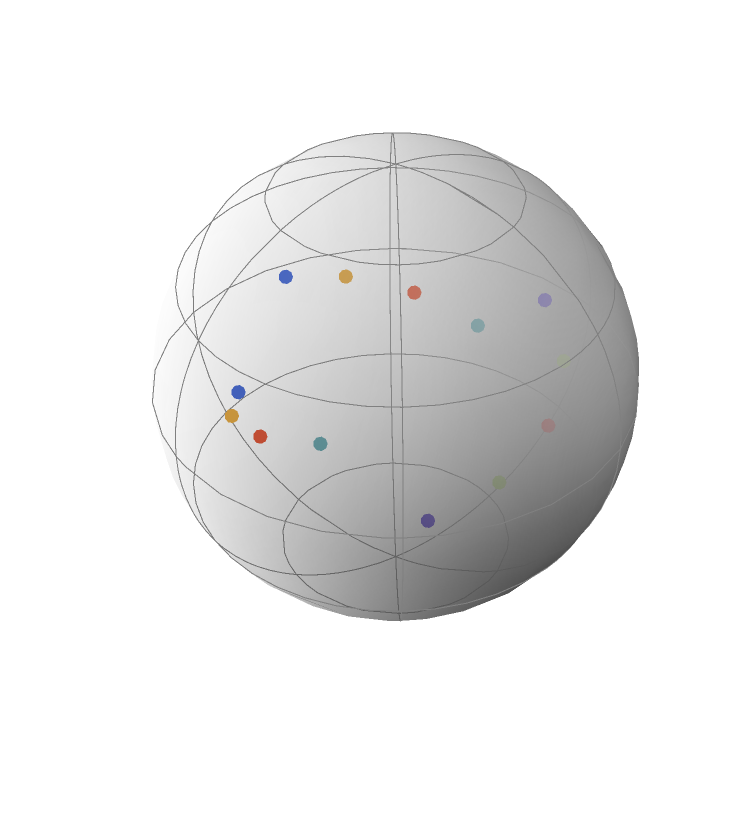}} \\ \hline
			\centered{$2000$ m}&\centered{\includegraphics[width=0.45\columnwidth]{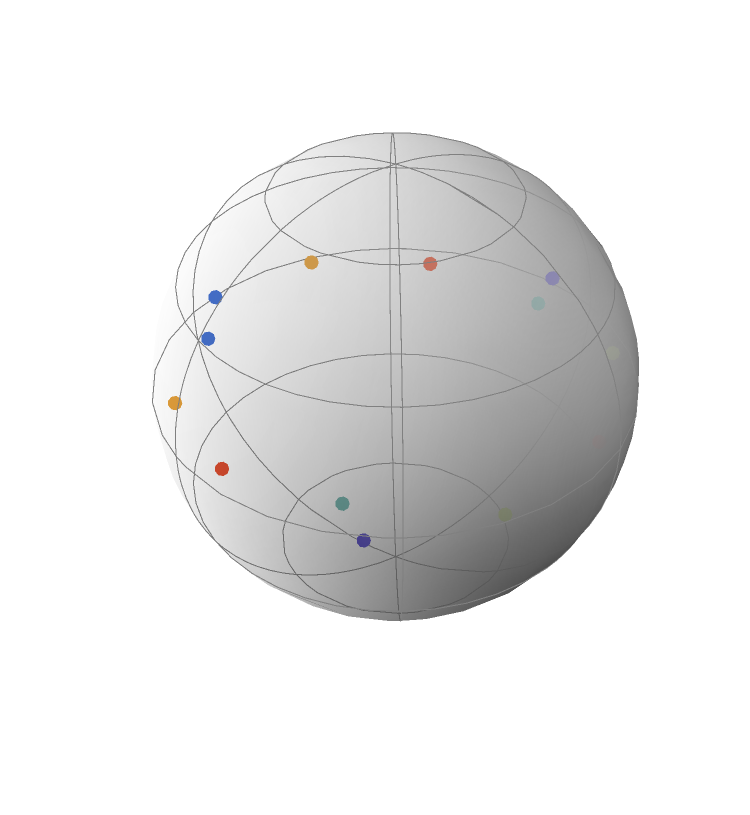}}\\
		\end{tabular}
		\caption{Graphical representation of the measurement operators in the Bloch ball for different values of the fiber length. The color in the temperature mapping represents the probability density of the measurement, i.e. $\mu(t)$ of the respective POVM element, where the most probable measurements are colored with red (the highest $\mu(t)$).}
		\label{spheres}
	\end{figure}

However, for some applications, it would be necessary to bear in mind that increasing the length of the fiber implies a boost of the attenuation, which involves a greater loss of photons and statistical uncertainty connected with measurements. In practice, one would need to possess a source of sufficient power to transmit photons over long distances. For a longer fiber, the power of the source should be turned up to compensate for the loss of photons due to greater attenuation. This is assumed in the present work since we consider the average number of photons per measurement as constant. Alternatively, it would be required to take into account the errors that may occur if the number of photons reaching the detectors drops down. In any case, before experimental applications an optimal balance between the gains and losses should be determined.

In \figureref{plotqubitphase}, one can notice that the shape of the function corresponding to $\sigma_D = 4$ ps appears intriguing. For fiber lengths such that $100 \text{m} < L < 325 \text{m}$, the function $\mathcal{F}_{min} (L)$ takes constant values, equal $\approx 0.5$. This particular property can be explained by visualizing the measurement operators \eqref{operator2} in the Bloch ball \figureref{spheres}. One can observe that for lengths belonging to the interval in question, the measurement operators arrange along a vertical line between the poles of the Bloch sphere. This proves that phase estimation, which is conducted for the states lying on the equator of the sphere, leads to the fidelity close to $0.5$. When the fiber length is increased, the representation of measurement operators in \figureref{spheres} gets inflated, which leads to better quality of phase estimation.

\subsection{Qutrits}

For qutrits the initial figures of merit have been gathered in \tabref{fig:qutrit}. One can instantly notice that if we consider the ideal scenario ($\sigma_D=0$ ps), the quantum tomography scheme does not lead to the perfect state recovery. Interestingly, even if there is no jitter we again observe an increase in the minimum fidelity provided we shift to the longer fiber.

The most interesting results have been obtained for $\sigma_D=1$ ps. Here we can observe the same kind of tendency as for qubits -- the quality of state reconstruction improves significantly if we consider the longer fiber. In particular, if $L=500$ m the state recovery based on the MLE appears to be very effective since the minimum fidelity equals $0.9009$, which is even more than in the scenario without any detector jitter.

Both QST methods appear to have trouble recovering an unknown density matrix for $\sigma_D=4$ ps. The figures of merit are weak for $L=200$ m and they only slightly improve if we use the longer fiber. Compared with qubits, one can conclude that in the case of qutrits the timing jitter has a more detrimental impact on the quality of quantum state reconstruction.

\begin{table}[h]
\centering
\bgroup
\def\arraystretch{2.1}
	\begin{tabular}{|c|c|c|c|c|c|}
\hline
		\multicolumn{2}{|c|}{\multirow{2}{*}{
				\backslashbox[22 mm]{\color{black}$\sigma_D$}{\color{black}$L\;$}}} & \multicolumn{2}{c|}{\centered{$200$ m}} & \multicolumn{2}{c|}{\centered{$500$ m}} \\ \cline{3-6} 
		\multicolumn{2}{|c|}{} & $\;\mathcal{F}_{min}\;$ & $\;\mathcal{D}_{max}\;$ & $\;\mathcal{F}_{min}\;$ & $\;\mathcal{D}_{max}\;$ \\ \hline
		\multirow{2}{*}{$0$ ps} &  LS  & $0.8058$  &$0.2859$  & $ 0.9180 $ &$0.1159$ \\ \cline{2-6} 
		& MLE & $0.8149$ & $0.2923$ & $0.8865$ & $0.1387$ \\ \hline

		\multirow{2}{*}{$1$ ps} & LS & $0.4817$ & $0.7084$ & $0.8490$ & $0.1548$  \\ \cline{2-6} 
		& MLE & $0.4824$ & $0.7077$ & $0.9009$ & $0.2250$ \\ \hline

		\multirow{2}{*}{$4$ ps} & LS & $0.2267$ & $0.8600$ & $0.2674$ & $0.8246$  \\ \cline{2-6} 
		& MLE & $0.1632$ & $0.9018$ & $0.2393$ & $0.8351$ \\
\hline
	\end{tabular}
\egroup
	\caption{Minimum fidelity and maximum trace distance in quantum tomography of qutrits. Figures computed numerically for different values of experimental parameters. The results were obtained after minimization over a sample of $9261$ qutrits.}
	\label{fig:qutrit}
\end{table}

\begin{figure}[h!]
	\centering
	\begin{tabular}{c}
	\centered{\includegraphics[width=1\columnwidth]{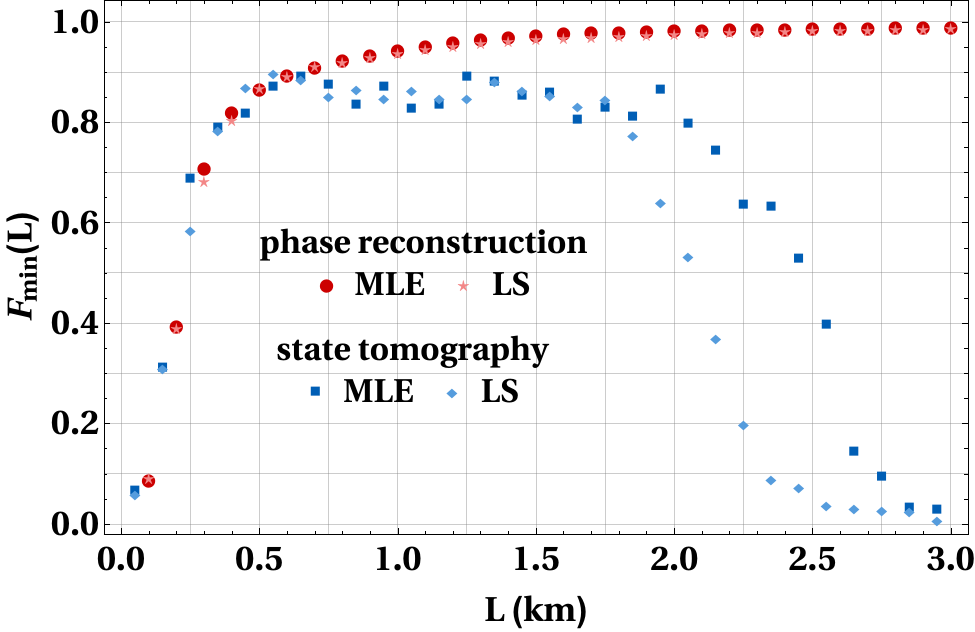}}
	\end{tabular}
	\caption{Plots of $\mathcal{F}_{min} (L)$ for complete state tomography and phase reconstruction of qutrits by both QST methods. Detector jitter is fixed $\sigma_D = 1$ ps.}
	\label{plotqutrits}
\end{figure}

In the case of qutrits, for any value of the detector jitter one gets better minimum fidelity for the longer fiber, though the extent of the enhancement differs. Since for $\sigma_D=1$ ps the scale of improvement is the greatest, we shall thoroughly investigate this instance. In \figureref{plotqutrits} there are the plots of $\mathcal{F}_{min} (L)$ for QST of qutrits (blue dots). Each point was obtained after minimization over a sample of $9261$ qutrits.

One can notice that as far as the LS method is concerned the minimum fidelity first increases along with the length of the fiber, then for any length such that $300 \,\text{m} \leq L \leq 1800 \,\text{m}$ we have $0.8 < \mathcal{F}_{min} (L) < 0.9$, which means that the quality of state reconstruction remains roughly stable for this spectrum of lengths. There is no clear maximum within this range. When $L$ exceeds $1800$ m the minimum fidelity quickly declines and around $L = 2000$ m drops below $0.5$.

The plot of $\mathcal{F}_{min} (L)$ by MLE presents very similar properties, though for two values of $L$, i.e. $L=500$ m and $L=550$ m, the minimum fidelity exceeds $0.9$. The minimum fidelity exceeds $0.8$ up to $L=2000$ m, when it starts to gradually decline towards zero.

Just as in the case of qubits, one may consider a specific subset of input qutrits which differ only in terms of the relative phase factors. If we denote by $\phi_{12}$ the relative phase between the states $\ket{1}$ and $\ket{2}$, then analogously we introduce $\phi_{13}$ (global phase is omitted), we shall consider qutrits of the form:
\begin{equation}\label{input3}
\rho_{in} (\phi_{12}, \phi_{13}) = \frac{1}{3} \begin{pmatrix}  1 & e^{- i \,\phi_{12}} & e ^{- i \,\phi_{13}} \\ e^{ i \,\phi_{12}} & 1 & e^{ i \,(\phi_{12}-\phi_{13})} \\ e ^{ i\, \phi_{13}} & e^{ i \,(\phi_{13}-\phi_{12})} & 1 \end{pmatrix}.
\end{equation}

Then, the input states are generated for a range of different values of the relative phases (a sample of $1681$ input qutrits was considered in simulations). Each input state was reconstructed, assuming no \textit{a priori} knowledge about the density matrix (i.e., we estimated the values of all $9$ parameters which characterize the density matrix of a qutrit). The results for phase reconstruction of qutrits are presented as the red dots in \figureref{plotqutrits}.

One can notice that the minimum fidelity for phase reconstruction of qutrits grows along with the length of the fiber, exceeds the value of $\mathcal{F}_{min} (L)$ for the complete state tomography and then asymptotically approaches $1$. The simulations have been carried on and the quality of phase estimation remains stable. This proves that phase recovery can be performed effectively even if we apply longer fibers.

\begin{figure}[h!]
	\centering
	\begin{tabular}{c}
		\centered{\includegraphics[width=0.95\columnwidth]{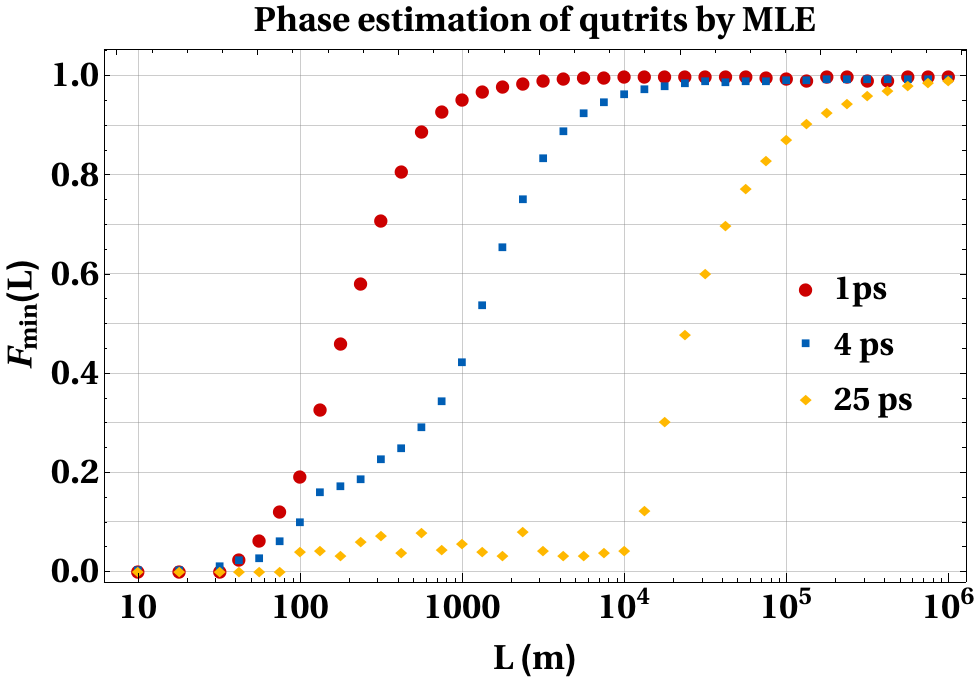}}
	\end{tabular}
	\caption{Plots of $\mathcal{F}_{min} (L)$ for phase estimation of qutrits performed by MLE in the case of $\sigma_D = 1$ ps (red), $\sigma_D = 4$ ps (blue) and $\sigma_D = 25$ ps (yellow). Fiber lengths are arranged in a logarithmic scale.}
	\label{plotqutritphase}
\end{figure}

When it comes to qutrits, one can also test whether the phase estimation is feasible within the abilities of current technology. We consider two higher values of the detector jitter: $\sigma_D = 4$ ps and $\sigma_D = 25$ ps, and compare the corresponding functions with $\sigma_D = 1$ ps. For such parameters, we perform phase estimation by our time-resolved measurement. The results are presented in \figureref{plotqutritphase}.

One can notice that for $\sigma_D = 25$ ps the quality of phase estimation remains very poor up to $L= 10^4$ m, when $\mathcal{F}_{min} (L)$ starts to grow rapidly to reach the value $\sim 1$ for $L = 10^6$ m. This means that even a realistic timing jitter, i.e. $\sigma_D = 25$ ps, can be compensated for by a sufficiently long fiber. If we consider $\sigma_D = 4$ ps, then on average we obtain better quality of phase estimation, but both functions converge at $L = 10^6$ m.

Generally, compared with qubits, it can be said that for $3-$level systems the framework is less efficient regarding the complete state tomography, but still by a proper choice of the fiber length one can ensure that the minimum fidelity shall come approximately to $0.9$. On the other hand, phase estimation of qutrits appears to be as effective as in the case of qubits but for higher values of the timing jitter one needs to consider longer fibers to compensate for the noise caused by the detector.

\section{Summary and outlook}

We have introduced a QST framework based on measurement operators defined in the time domain. In our scenario, the measured photon counts are distorted by the detector timing jitter which is a major source of errors in time-bin encoding. By the proper selection of the figures of merit we could evaluate the performance of the tomographic technique. As the main result, we have presented and discussed the minimum fidelity as a function of the length of the fiber. The results contribute to the area of quantum state estimation since we have indicated the optimal conditions for density matrix reconstruction.

In particular, we have demonstrated that phase recovery is feasible by our tomographic technique for both qubits and qutrits. Even if we assume realistic values of the detector jitter, we can perform high-quality phase estimation with the time-resolved single-photon counting. Therefore, the method is in accordance with current technological capabilities. In the future, the framework shall be tested on multilevel systems and applied in experiments.

\section*{Acknowledgements}

K.~S.-K. was supported by the National Science Centre, Poland (NCN) (Grant No.~2016/23/N/ST2/02133). A.~C. acknowledges the support from the Foundation for Polish Science (FNP) (project First Team co-financed by the European Union under the European Regional Development Fund). P.~K. acknowledges the support from the National Science Centre, Poland (NCN) (Grant No.~2016/23/D/ST2/02064). All the authors acknowledge support by the National Laboratory of Atomic, Molecular and Optical Physics, Torun, Poland.

\end{document}